\documentstyle[sprocl,pictex]{article}

\def\Journal#1#2#3#4{{#1} {\bf #2}, #3 (#4)}

\def\NPB{{\em Nucl. Phys.} B}
\def\PLB{{\em Phys. Lett.}  B}
\def\PRL{\em Phys. Rev. Lett.}

\def\be{\begin{equation}}
\def\ee{\end{equation}}
\def\bea{\begin{eqnarray}}
\def\eea{\end{eqnarray}}

\begin{document}

\title{SUPERSTRING AXION, GAUGINO CONDENSATION AND DISCRETE SYMMETRIES
\footnote{Talk presented at PASCOS-98, Northeastern University, Boston,
March 22--29, 1998.}}

\author{JIHN E. KIM}

\address{Department of Physics and Center for Theoretical
Physics\\
Seoul National University, Seoul 151-742, Korea, and\\
School of Physics, Korea Institute for Advanced Study,\\
207-43 Cheongryangri-dong, Seoul 130-012, Korea}

\maketitle\abstracts{Here I present the effect of gaugino condensation
and discrete symmetries to the model-independent axion potential,
collaborated with Georgi and Nilles \cite{gkn}. It is
shown that with appropriate discrete symmetries the model-independent
axion can solve the strong CP problem}

\section{Introduction}

For a confining gauge group, we have a physical $\theta$ parameter.
For QCD, this parameter $\bar\theta$ is known to be extremely
small,
\begin{equation}
|\bar\theta|<10^{-9}
\end{equation}
from the upper bound of neutron electric dipole moment. Thus $\bar\theta$
is a very small parameter of the standard model. This smallness of the
parameter is one of the parameter problems, usually known as the
strong CP problem. On the other hand, from instanton calculus, the
$\bar\theta$-dependence of $V$ is such that it is minimum at 
$\bar\theta=0$. $V$ is periodic with period of $2\pi$.

%
%
\begin{figure}[htb]
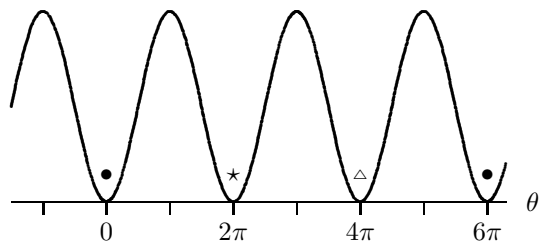

$$\beginpicture
\setcoordinatesystem units <72pt,36pt> point at 0 0
\setplotarea x from -0.500 to 2.100, y from 0.000 to 2.000
\axis bottom 
      ticks out
      width <0.5pt> length <5.0pt>
      withvalues {} {$0$} {} {$2\pi$} {} {$4\pi$} {} {$6\pi$} {} /
      at -0.333 0.000 0.333 0.666 1.000 1.333 1.666 2.000 /
/
\put {$\theta$} <10pt,0pt> at 2.100 0.000
\setplotsymbol ({\normalsize.})
\setsolid
\setquadratic
\plot
-0.500  1.000 -0.450  1.454 -0.400  1.809 -0.350  1.988 -0.300  1.951
-0.250  1.707 -0.200  1.309 -0.150  0.844 -0.100  0.412 -0.050  0.109
 0.000  0.000  0.050  0.109  0.100  0.412  0.150  0.844  0.200  1.309
 0.250  1.707  0.300  1.951  0.350  1.988  0.400  1.809  0.450  1.454
 0.500  1.000  0.550  0.546  0.600  0.191  0.650  0.012  0.700  0.049
 0.750  0.293  0.800  0.691  0.850  1.156  0.900  1.588  0.950  1.891
 1.000  2.000  1.050  1.891  1.100  1.588  1.150  1.156  1.200  0.691
 1.250  0.293  1.300  0.049  1.350  0.012  1.400  0.191  1.450  0.546
 1.500  1.000  1.550  1.454  1.600  1.809  1.650  1.988  1.700  1.951
 1.750  1.707  1.800  1.309  1.850  0.844  1.900  0.412  1.950  0.109
 2.000  0.000  2.050  0.109  2.100  0.412
/
\put {$\bullet$} <0pt,10pt> at 0.000 0.000
\put {$\star$} <0pt,10pt> at 0.667 0.000
\put {\tiny$\triangle$} <0pt,10pt> at 1.333 0.000
\put {$\bullet$} <0pt,10pt> at 2.000 0.000
\endpicture$$
\caption{V[$\theta$] versus $\theta$. In axionic models, $\theta=a/F_a$.}
\label{fig.1}
\end{figure}

\noindent
If $\bar\theta$ is treated as a coupling, then a theory with any
$\theta$ will become a good theory, but 
the bound~(1) excludes the most regions of $\bar\theta$, which is
the strong CP problem. It is most elegantly solved in the axion
framework. If $\bar\theta$ is a dynamical field, i.e. if it
appears with the kinetic energy term, then $\bar\theta$ slides
down the hill of $V[\bar\theta]$, and will eventually settle at
$\bar\theta=0$, which is the axion solution \cite{review}.

In axion models, therefore, one identifies $\bar\theta$ as an axion $a$, 
\begin{equation} 
\bar\theta={a\over F_a}
\end{equation}
Note that the axion $a$ does not have any potential except that coming
from the $\bar\theta F\tilde F$ term. Otherwise, the mechanism does
not work. As given in Eq.~(2), the axion model introduces a mechanism to
introduce the so-called axion decay constant $F_a$. It can arise from
a spontaneous symmetry breaking scale, or from the gravitational
scale, or from a compositeness scale. Thus the axion can be attractive 
to those who emphasize one of these points as the most fundamental
aspect of the particle theory. 

There are many interesting physical phenomena due to the existence of
axion: domain walls, axionic strings, cosmic axion energy density,
galaxy formation, stellar energy loss, etc. From these, studies, we
can restrict the range of the axion decay constant to
$10^9\ {\rm GeV}<F_a<10^{12-13}\ {\rm GeV}$,
where the upper bound is slightly extended since heavy particle
decays can raise the original bound estimate. Now the cavity
experiments are going on to detect the galactic axions corresponding
to the upper bound of $F_a$ \cite{kim98}.

>From now on, I will try to discuss the resurrection of the axion 
in the string theory.

\section{Simple Solutions of the Strong CP Problem}

There are two examples in which the strong CP problem is 
automatically solved. One is the axion solution, and the other is a
massless u-quark solution.  Since our discussion on the 
model-independent axion solution relies on these aspect, we briefly
review the ideas of these two. 

\subsection{Heavy Quark Axion}

The simplest axion model is the heavy quark axion in
which only $\bar\theta$ is introduced in addition to the
standard model fields below the axion scale $F_a$ \cite{kim79}. 
The Lagrangian is
$$
{\cal L}=\sigma\bar Q_RQ_L+{\rm h.c.}-V
$$
where we suppressed couplings and $V$ is assumed to respect
the PQ symmetry $U(1)_A$,
\begin{equation}
Q_L\rightarrow e^{-i{\alpha\over 2}}Q_L,\ Q_R\rightarrow
e^{i{\alpha\over 2}}Q_R,\ \sigma\rightarrow e^{i\alpha}
\sigma,\ \theta\rightarrow\theta-\alpha
\end{equation}
For a nonzero VEV $<\sigma>=F_a/\sqrt{2}$, $Q$ obtains a mass
at scale $F_a$, the radial component $\rho$ (Higgs--type field) 
of $\sigma$ obtains a mass at scale $F_a$, and at low energy 
there remains only the axion $a$. Thus from the kinetic 
term $D_\mu\sigma^*D^\mu\sigma$, we obtain $(1/2)\partial_\mu
a\partial^\mu a$ where $\sigma=([F_a+\rho]/\sqrt{2})e^{ia/F_a}$.

Thus, below the scale $F_a$, the light fields are gluons and
$a$ (plus the other SM fields). The relevant part of the
Lagrangian respecting the symmetry (4) (i.e. with $a
\rightarrow a+\alpha F_a$) is
\begin{equation}
{\cal L}={1\over 2}(\partial_\mu a)^2+{\rm (derivative\ term\ 
of\ } a) +(\theta+{a\over F_a}){1\over 32\pi^2}F^a_{\mu\nu}
\tilde F^{a\mu\nu}
\end{equation} 
Note that we created the needed 
$F\tilde F$ coupling minimally. Usually,
$a$ is redefined as $a-\theta F_a$ so that the coefficient of
$F\tilde F$ is $a/F_a\equiv\bar\theta$. Thus, the above
effective Lagrangian is seen to be invariant under the symmetry
transformation Eq.~(4).

\subsection{Massless u-quark Solution}

To see the $\theta$-independence of the effective Lagrangian of
QCD below the chiral symmetry breaking scale with the massless
u-quark, let us consider the
one-flavor QCD first with a mass parameter $m_u$,
$$
{\cal L}=-m_u\bar u_Ru_L+{\rm h.c.}
$$
which possess the following hypothetical symmetry
\begin{equation}
u_L\rightarrow e^{i\alpha}u_L,\ \bar u_R\rightarrow e^{i\alpha}
\bar u_R,\ m_u\rightarrow e^{-2i\alpha}m_u,\ 
\theta\rightarrow \theta+2\alpha 
\end{equation}
Since $m_u$ is endowed with a transformation even
though it is not a symmetry, Eq.~(5) is
useful to trace the $m_u$ dependence in the effective theory
below the quark condensation scale $<\bar uu>\propto v^3e^{i\eta/v}$,
\begin{eqnarray}
&V={1\over 2}m_u\Lambda^3e^{i\theta}-{1\over 2}\lambda_1
\Lambda v^3e^{i(\eta/v-\theta)}
-{1\over 2}\lambda_2m_uv^3e^{i\eta/v}\nonumber\\
&+\lambda_3m_u^2\Lambda^2e^{2i\theta}
+\lambda_4{v^6\over\Lambda^2}
e^{2i(\eta/v-\theta)}+\cdots+{\rm h.c.}\nonumber
\end{eqnarray}
where the strong interaction scale $\Lambda$ is inserted to make
the dimension appropriate.
Note that, for $m_u=0$, $\eta-F_a\theta$ can be redefined as a new
$\eta$, removing the $\theta$ dependence. Thus $\theta$ is
unphysical in a massless $u$-quark theory, solving the strong
CP problem. 

Since the interactions at the gravitational scale may violate
global symmetries, one can consider the following nonrenormalizable
interactions for the massless u-quark case,
\begin{equation}
{1\over M_P^2}\bar u\bar u uu e^{-2i\theta},\ {1\over M_P}
\bar uu\sigma\sigma e^{-i\theta} 
\end{equation}
where $\sigma$ is a singlet scalar field. If only the first term
is the only allowed nonrenormalizable interaction, $\theta$ is
shifted by a tiny amount, $10^{-38}$; thus the massless u-quark
solution is still valid. However, if the second term is allowed
with nonvanishing VEV of $\sigma$, the VEV must be bounded to be less
than $10^4$ GeV to have phenomenologically allowable $\theta$,
given in Eq.~(1). Thus, the massless u-quark idea is not automatically
solving the strong CP problem with gravitational interactions.

It is interesting to see how one can obtain the axion
mass from the above symmetry argument.
At the minimum of the potential, the $a$--$\eta$
mass matrix for $m_u\ne 0$ is
$$¤
M^2=\left(\matrix{&\lambda\Lambda v+\lambda' mv, 
&-\lambda\Lambda v^2/F_a\cr
&-\lambda\Lambda v^2/F_a, 
&-{m\Lambda^3\over F_a^2}+{\lambda\Lambda v^3\over F_a^2}\cr}\right)
$$
Diagonalizing the above mass matrix for $F_a\gg$ (other mass
parameters), we obtain for vacuum at $\theta=0$
$$
m_a^2={m_u\Lambda\over F_a^2}\left({\lambda\lambda' v^4\over
\lambda\Lambda v+\lambda' m_uv}-\Lambda^2\right),\  
m_\eta^2=(\lambda\Lambda+\lambda' m)v
$$
which shows the essential features of the axion mass: it is
suppressed by $F_a$, multiplied by $m_q$, and the rest of
condensation parameters. If the above mass squared is negative,
we chose a wrong vacuum and choose $\theta=\pi$ 
instead as the vacuum.

For a realistic axion mass, however, we consider one family QCD
$$
{\cal L}=-m_u\bar uu-m_d\bar dd
$$
which possesses the fictitious $U(1)_u\times U(1)_d$ symmetry,
\begin{eqnarray}
&u_L\rightarrow e^{i\alpha}u_L,\ d_L\rightarrow e^{i\beta}d_L,
\ m_u\rightarrow e^{-2i\alpha}m_u,\nonumber\\
&m_d\rightarrow e^{-2i\beta}m_d,\ \theta\rightarrow \theta
+2(\alpha+\beta)
\end{eqnarray}
Following the same procedure, we obtain
\begin{equation}
m_a={m_{\pi^0}F_\pi\over F_a}{\sqrt{Z}\over 1+Z}
\end{equation}
where $Z=m_u/m_d$. The above formula is valid for the
KSVZ model. For the PQWW and DFSZ models,
one needs extra consideration, for removing the 
longitudinal component of $Z^0$. In the limit of
$F_a\gg$ (other mass parameters), i.e. in the DFSZ model,  
the above formula is also valid.

\section{Superstring Axion}

The standard introduction of axion through spontaneous breaking
of $U(1)_A$ global symmetry is {\it ad hoc}. The reason is that
there exist so many ways to introduce the symmetry.

There exists another very fundamental way to introduce
the axion. It is in the string theory. Furthermore,
the axion {\it must be present} in string theory for the automatic
strong CP solution in string models. Nevertheless, the
route to the axion solution in string models is not
always present, and therefore let us first see what are the
problems along this line and then present a possible
route.

Ten dimensional string models contain massless 
bosons $G_{MN}$ ($MN$ symmetric), $B_{MN}$ ($MN$ antisymmetric), 
and $\phi$, where $M,N$ run through indices 
0,1,$\cdots$,9. Our interest here
is the antisymmetric tensor field $B_{MN}$ which contains
two kinds of axions: model-independent axion (MIa) \cite{witten} and 
model-dependent axions \cite{witten1}. 

The MIa is basically $B_{\mu\nu}$ where $\mu,\nu$ is the
4D indices $0,1,2,3$. The dual of the field strength is
defined as the derivative of MIa $a$,
\begin{equation} 
\partial^\sigma a\sim \epsilon^{\mu\nu\rho\sigma}H_{\mu\nu\rho},
H_{\mu\nu\rho}\sim \epsilon_{\mu\nu\rho\sigma}\partial^\sigma a
\end{equation}
The question is why we interpret this as an axion. It
is due to Green, Schwarz, and Witten \cite{gs,witten}.
The gauge invariant field strength $H$ of $B$ is 
$H=dB-\omega^0_{3Y}+\omega^0_{3L}$ with the Yang-Mills Chern-Simmons term
$\omega^0_{3Y}={\rm tr}(AF-A^3/3)$ and the Lorentz 
Chern-Simmons term $\omega^0_{3L}={\rm tr}(\omega R-\omega^3/3)$.
These satisfy $d\omega^0_{3Y}={\rm tr}F^2$ and $d\omega^0_{3L}
={\rm tr}R^2$. Therefore,
\begin{equation}
dH=-{\rm tr}F^2+{\rm tr}R^2
\end{equation}
Note also that one had to introduce a nontrivial gauge
transformation property of $B$. Then gauge anomaly is
completely cancelled by introducing the so-called
Green-Schwarz term \cite{gs}, 
\begin{equation}
S_{GS}\propto \int (B{\rm tr}F^4+\cdots)
\end{equation}
which contains the coupling of the form
\begin{eqnarray}
&\epsilon_{ijKLMNOPQR}B_{ij}F^{KL}F^{MN}F^{OP}F^{QR}\sim\nonumber\\
&B_{ij}(\epsilon_{\mu\nu\rho\sigma}F^{\mu\nu}F^{\rho\sigma})
<F_{kl}><F_{pq}>\epsilon_{ijklpq}\nonumber
\end{eqnarray}
where we note the Minkowski indices $\mu,\nu,\cdots$ and the
internal space indices $i,j,\cdots$. The nonvanishing VEV
$<F>$ gives $a'F\tilde F$ coupling at tree level. Thus we
are tempted to interpret $a'$ as an axion, and it was called
model-dependent axion \cite{witten1,kiwoon}. But we have to
check that there is no dangerous potential term involving
$a'$, and indeed it has been shown that world-sheet instanton 
contribution
$$
i\int_{\Sigma_J}d^2z B_I\omega^I_{i\bar j}(\partial X^i\bar
\partial X^{\bar j}-\bar\partial X^i\partial X^{\bar j})
=2\alpha' B_J
$$
gives $a'(=B_I)$ dependent superpotential \cite{wen},
removing $a'$ as a useful degree for relaxing a
vacuum angle. In the above equation, $\alpha'$
is the string tension and $\omega^I_{i\bar j}$
represents the topology of the internal space,
\begin{equation}
B=B_{\mu\nu}dx^\mu dx^\nu+B_I\omega^I_{i\bar j}
dz^id\bar z^{\bar j}
\end{equation}
But $B_{\mu\nu}$ is still good since it does not get
a contribution from the stringy world-sheet instanton
effect. Eq. (10) implies
\begin{equation}
\Box a=-{1\over M}({\rm Tr}F_{\mu\nu}\tilde
F^{\mu\nu}-{\rm Tr}R_{\mu\nu}\tilde R^{\mu\nu})
\end{equation}
obtained from an effective Lagrangian of the form
\begin{equation}
{\cal L}={1\over 2}(\partial_\mu a)^2-{a\over 16\pi^2F_a}({\rm Tr}
F_{\mu\nu}\tilde F^{\mu\nu}-\cdots)
\end{equation}
Thus $a$ is the axion (MIa). Any string models have this
MIa and its decay constant is of order
$10^{16}$ GeV \cite{fmia}. At this point, we comment that there are two
serious problems of the superstring axion:

\noindent {\it (A) The axion decay constant problem}--It is known
that the decay constant of MIa is of order $10^{16}$ GeV
\cite{fmia} which is far above the cosmological upper bound 
of $F_a$. A large $F_a$ can be reconciled with
cosmological energy density if a sufficient number of radiation 
are added below 1 GeV of the universe temperature,
but then it is hopeless to detect the cosmic axion by cavity 
detectors. Therefore, $F_a$ is better to be lowered to
around $10^{12}$ GeV.  

\noindent {\it (B) The hidden sector problem}--It is a
general belief here that a hidden sector confining force,
e.g. $SU(N)$, is needed for supersymmetry breaking at
$10^{12}\sim 10^{13}$ GeV. If so, MIa gets mass also due
to the $a_{MI} F'\tilde F'$ coupling where $F'$ is the
field strength of the hidden sector confining gauge field,
and we expect $m_a\simeq \Lambda_h^2/F_a$ which is obviously
too large to bring down MIa to low energy scale for the
solution of the strong CP problem. For example, the axion
gets potential both from the hidden sector scale $\Lambda_h$
and the QCD scale $\Lambda_{QCD}$ (if there is no matter)
in the follwing way,
$$
-\Lambda_{QCD}^4\cos({a_{MI}\over F_a}+\theta^0)-\Lambda_h^4
\cos({a_{MI}\over F_a}+\theta_h^0)
$$
where we added two terms with independent phases $\theta^0,
\theta_h^0$ which arise at the string scale when CP is
broken. Because $\Lambda_h\gg\Lambda_{QCD}$, the vacuum
chooses ${a_{MI}\over F_a}+\theta_h^0=0$, i.e. $<a_{MI}>\simeq
-\theta_h^0 F_a$, implying $\bar\theta\simeq\theta^0-\theta_h^0$
which is not zero in general. If we want to settle both $\theta_h$
and $\bar\theta$ at zero, then we need two independent axions.
However, it is known that only MIa is available at string induced 
low energy physics. Therefore, we say that there is the hidden
sector problem in the MIa phenomenology.

The above two problems have to be resolved if the string theory
render an acceptable low energy standard model and also if
the axion solution has a profound root in the fundamental theory
of the universe. It turns out that it is very difficult to
achieve. Only in a limited case, it may be possible to
find a possible route toward a solution.

\section{Anomalous U(1)}

In anomalous $U(1)$ gauge models \cite{anom}, 
the Green-Schwarz term contains
\begin{equation}
\epsilon_{MNOPQRSTUV}B^{MN}\cdot {\rm Tr}F^{OP}
\cdot <F^{QR}><F^{ST}><F^{UV}>
\end{equation}
which introduces the coupling $M_c(\partial^\mu a_{MI})A_\mu$.
Namely, the MIa becomes the longitudinal degree of freedom of
$A_\mu$. Thus, the $U(1)_A$ gauge boson becomes massive
and $a_{MI}$ is removed at low energy. Below the scale
$M_c$, then there exists a global symmetry \cite{ami,gkn}. 

Superstring models also need an extra
hidden sector confining force. Then, even if
the MIa is present, it obtains a dominant contribution to the mass 
from the hidden sector instanton effects. One can make the
contribution to the MIa mass absent if there is a massless
hidden colored fermion. It is similar to the massless u-quark
solution of the strong CP problem. 
The first obvious choice is the theory
of a massless hidden sector gaugino without a hidden matter.
But the hidden sector gaugino is NOT massless. Nevertheless,
the final hidden sector gaugino mass is not introduced by hand, but
it arises from the condensation of the hidden sector gauginos,
thus the contribution to the MIa potential is absent in this
limit. However, the string (or gravitational) theory does not
preserve any global symmetry, thus there must be interactions
violating the $R$ symmetry.\footnote{With the massless gaugino
with only renormalizable gaugino self interactions, there is an
$R$ symmetry.} It is similar to the interactions~(6) in massless
u-quark model.  

Thus we introduce an anomalous $U(1)$.  The hidden sector gluino is
massless in supersymmetric cases. With supersymmetry breaking,
the gluino will obtain mass eventually; but this case is different
from nonzero mass u-quark case since the hidden sector gluino
obtains mass by the hidden sector gluino condensation. So if the
R-symmetry is not explicitly broken, there is no contribution to
the potential of MIa from the hidden sector. However, the
gravitational interactions violate the global symmetry. For
example, with $SU(N)$ hidden gauge group, we expect
\begin{equation}  
 V_{\rm eff}={1\over 2}\bar\lambda\bar\lambda\lambda\lambda+
\epsilon_2\bar\lambda\bar\lambda\lambda\lambda\lambda\lambda+
(\lambda\lambda)^Ne^{-i\bar\theta}+{\rm h.c.}+\cdots 
\end{equation}
Endowing the following fictitious symmetry, $\lambda\rightarrow 
e^{i\alpha}\lambda,\theta_h\rightarrow\theta_h+2l(G)\alpha$
(where $l({\rm SU(N)})=N$),
and $\epsilon_2\rightarrow e^{-2i\alpha}\epsilon_2$,
the effective potential contains
$$
\epsilon_2\left({v\over M_P}\right)^5v^4e^{i\eta/v}
+\left({v\over\Lambda_h}\right)^{3N}\Lambda_h^4
e^{i(\eta N/v-\theta_h)}
$$
Note that, if $\epsilon_2=0$, then we do not have a $\theta_h$
dependence. The $\cdots$ in Eq.~(16) contains higher order terms
beyond the $\epsilon_2$ term; but if $U(1)_R$ were exact then
there would be no $\theta_h$ dependence of $V_{\rm eff}$, 

%
%
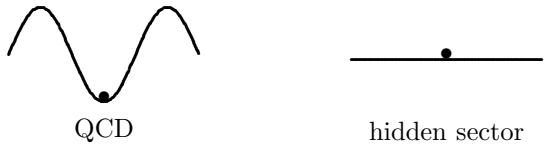
\begin{figure}[htb]
$$\beginpicture
\setcoordinatesystem units <72pt,18pt> point at 0 0
\setplotarea x from -0.700 to 2.100, y from 0.000 to 2.1000
\setplotsymbol ({\normalsize.})
\setsolid
\setquadratic
\put {QCD} [c] at -0.200 -0.600 
\put {{\rm hidden sector}} [c] at 1.600 -0.600 
\plot
-0.700  1.000 -0.650  1.454 -0.600  1.809 -0.550  1.988 -0.500  1.951
-0.450  1.707 -0.400  1.309 -0.350  0.844 -0.300  0.412 -0.250  0.109
-0.200  0.000 -0.150  0.109 -0.100  0.412 -0.050  0.844  0.000  1.309
 0.050  1.707  0.100  1.951  0.150  1.988  0.200  1.809  0.250  1.454
 0.300  1.000  
/
\put {$\bullet$} <0pt,2pt> at -0.200 0.000
\setlinear 
\plot
1.100 0.900 2.100 0.900
/
\put {$\bullet$} <0pt,2pt> at 1.600 0.900
\endpicture$$
\end{figure}

\centerline{\bf Discrete Symmetry $Z_N\subset U(1)_R$}
\vskip 0.3cm
But the above figure cannot be realized since $U(1)_R$ is not exact.
Therefore, the best we can hope is that only a discrete subgroup
of $U(1)_R$ is exact. Starting from a massless hidden sector gluino,
an obvious choice is a subgroup $Z_N$ of $U(1)_R$. Depending on
the compactification schemes, only a limited class of terms are
allowed in the superpotential. One such example is $W\sim T^{18n+12}$
$n=$ (integer) in some orbifold compactification where 
$T$ is the twisted sector fields \cite{ibanez}. Then a possible
unbroken discrete subgroup is $Z_{3n}$.

It is straightforward to estimate the hidden-sector contribution
to the MIa potential with an unbroken $Z_N$ which is shown below,

\vskip 0.2cm
\centerline{Table 1. The $\theta_h$ dependence of potential in 
GeV$^4$ units for $\Lambda_h=10^{12}-10^{13}$ GeV}

\begin{center}
\begin{tabular}{|ccccc|}
\hline
$Z_N$ &\ & $\epsilon_n$ &\ & $V$\\
\hline
$N$=2 &\ &\ \ \ $n$=2 &\ & $\ \ \sim 10^{13}-10^{25}$\\
$N$=3 &\ &\ \ \ $n$=4 &\ & $\ \ \sim 10^{-29}-10^{-8}$\\
$N$=4 &\ &\ \ \ $n$=3 &\ & $\ \ \sim 10^{-8}-10^{10}$\\
$N$=5 &\ &\ \ \ $n$=6 &\ & $\ \ \sim 10^{-50}-10^{-26}$\\
\hline
\end{tabular}
\end{center}
\vskip 0.3cm

Thus, for $Z_N$ with $N=3,5,6,\cdots$ the hidden sector
contribution to the MIa potential is negligible and MIa
acts as the invisible axion of the standard model.
Indeed, there are possibilities of realizing these $Z_N$
subgroups in string models.

\section{CONCLUSION}

The $\bar\theta$ parameter problem in the standard model must
be resolved in an ultimate theory if it exists. The axion
solution is the most attractive one. Usually, it is
believed that the gravitational interaction breaks all
global symmetries, and the basis for the axion is shaken.
The same comment applies to a massless quark solution.
However, the string models contain the MIa which can
be a good candidate for the nonlinearly realized global
symmetry. Nevertheless, we must pass through the hurdle of
the hidden sector physics.

The MIa gets
contributions to its potential from QCD and the hidden sector
as

%
%
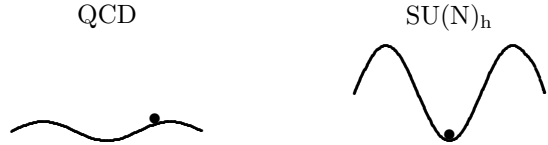
\begin{figure}[htb]
$$\beginpicture
\setcoordinatesystem units <72pt,18pt> point at 0 0
\setplotarea x from -0.700 to 2.100, y from 0.000 to 1.500
\setplotsymbol ({\normalsize.})
\setsolid
\setquadratic
\put {QCD} [c] at -0.200 2.600 
\put {SU(N)$_{\rm h}$} [c] at 1.600 2.600 
\plot
-0.700  0.2   -0.650  0.291 -0.600  0.362 -0.550  0.398 -0.500  0.390 
-0.450  0.341 -0.400  0.262 -0.350  0.169 -0.300  0.082 -0.250  0.022
-0.200  0.000 -0.150  0.022 -0.100  0.082 -0.050  0.169  0.000  0.262
 0.050  0.341  0.100  0.390  0.150  0.398  0.200  0.362  0.250  0.291
 0.300  0.2  
/
\put {$\bullet$} <0pt,2pt> at  0.050 0.341 
\plot
 1.1    1.000  1.15   1.454  1.200  1.809  1.250  1.988  1.300  1.951 
 1.350  1.707  1.400  1.309  1.450  0.844  1.500  0.412  1.550  0.109
 1.600  0.000  1.650  0.109  1.700  0.412  1.750  0.844  1.800  1.309
 1.850  1.707  1.900  1.951  1.950  1.988  2.000  1.809  2.050  1.545
 2.1    1.000    
/
\put {$\bullet$} <0pt,2pt> at 1.600 0.000
\endpicture$$
\end{figure}

\noindent
But with an anomalous $U(1)$ and $U(1)_R$ symmetries it gets as
%
%
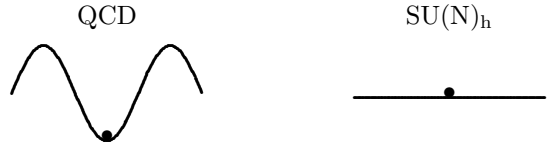
\begin{figure}[htb]
$$\beginpicture
\setcoordinatesystem units <72pt,18pt> point at 0 0
\setplotarea x from -0.700 to 2.100, y from 0.000 to 1.000
\setplotsymbol ({\normalsize.})
\setsolid
\setquadratic
\put {QCD} [c] at -0.200 2.600 
\put {SU(N)$_{\rm h}$} [c] at 1.600 2.600
\plot
-0.700  1.000 -0.650  1.454 -0.600  1.809 -0.550  1.988 -0.500  1.951
-0.450  1.707 -0.400  1.309 -0.350  0.844 -0.300  0.412 -0.250  0.109
-0.200  0.000 -0.150  0.109 -0.100  0.412 -0.050  0.844  0.000  1.309
 0.050  1.707  0.100  1.951  0.150  1.988  0.200  1.809  0.250  1.454
 0.300  1.000  
/
\put {$\bullet$} <0pt,2pt> at -0.200 0.000
\setlinear 
\plot
1.100 0.900 2.100 0.900
/
\put {$\bullet$} <0pt,2pt> at 1.600 0.900
\endpicture$$
\end{figure}

\noindent
Gravitational interaction may break $U(1)_R$ and we expect the
%
%
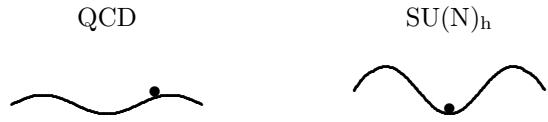
\begin{figure}[htb]
$$\beginpicture
\setcoordinatesystem units <72pt,18pt> point at 0 0
\setplotarea x from -0.700 to 2.100, y from 0.000 to 1.000
\setplotsymbol ({\normalsize.})
\setsolid
\setquadratic
\put {QCD} [c] at -0.200 2.000 
\put {SU(N)$_{\rm h}$} [c] at 1.600 2.000 
\plot
-0.700  0.2   -0.650  0.291 -0.600  0.362 -0.550  0.398 -0.500  0.390 
-0.450  0.341 -0.400  0.262 -0.350  0.169 -0.300  0.082 -0.250  0.022
-0.200  0.000 -0.150  0.022 -0.100  0.082 -0.050  0.169  0.000  0.262
 0.050  0.341  0.100  0.390  0.150  0.398  0.200  0.362  0.250  0.291
 0.300  0.2  
/
\put {$\bullet$} <0pt,2pt> at  0.050 0.341 
\plot
 1.1    0.500  1.15   0.772  1.200  0.904  1.250  0.994  1.300  0.975 
 1.350  0.850  1.400  0.654  1.450  0.422  1.500  0.206  1.550  0.054 
 1.600  0.000  1.650  0.054  1.700  0.206  1.750  0.422  1.800  0.654
 1.850  0.85   1.900  0.975  1.950  0.994  2.000  0.904  2.050  0.772 
 2.1    0.500     
/
\put {$\bullet$} <0pt,2pt> at 1.600 0.000
\endpicture$$
\end{figure}

\noindent
Even if the $U(1)_R$ symmetry is broken, the hidden sector contribution
is not as large as $\Lambda_h^4$, because the hidden sector gluino obtains
mass through the gluino condensation itself. This is a different point
from the massive u-quark case where u-quark mass is given as a parameter
in QCD. 

But if $Z_{3,5}, \cdots$ subgroup of $U(1)_R$ is unbroken
the hidden sector contribution is negligible 
%
%
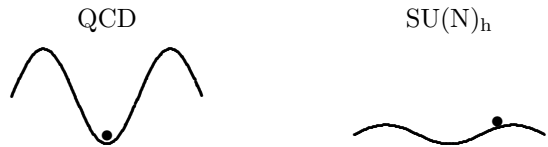
\begin{figure}[htb]
$$\beginpicture
\setcoordinatesystem units <72pt,18pt> point at 0 0
\setplotarea x from -0.700 to 2.100, y from 0.000 to 1.000
\setplotsymbol ({\normalsize.})
\setsolid
\setquadratic
\put {QCD} [c] at -0.200 2.600 
\put {SU(N)$_{\rm h}$} [c] at 1.600 2.600 
\plot
-0.700  1.000 -0.650  1.454 -0.600  1.809 -0.550  1.988 -0.500  1.951
-0.450  1.707 -0.400  1.309 -0.350  0.844 -0.300  0.412 -0.250  0.109
-0.200  0.000 -0.150  0.109 -0.100  0.412 -0.050  0.844  0.000  1.309
 0.050  1.707  0.100  1.951  0.150  1.988  0.200  1.809  0.250  1.454
 0.300  1.000  
/
\put {$\bullet$} <0pt,3pt> at -0.200 0.000
\plot
 1.1    0.2    1.15   0.291  1.200  0.362  1.250  0.398  1.300  0.390
 1.350  0.341  1.400  0.262  1.450  0.169  1.500  0.082  1.550  0.022
 1.600  0.000  1.650  0.022  1.700  0.082  1.750  0.169  1.800  0.262
 1.850  0.341  1.900  0.390  1.950  0.398  2.000  0.362  2.050  0.291
 2.1    0.2    
/
\put {$\bullet$} <0pt,2pt> at 1.850 0.341
\endpicture$$
\end{figure}

Thus superstring models can include the invisible axion
needed for the strong CP solution. 

\section*{Acknowledgments}
This work is supported by Distinguished Scholar Exchange Program
of Korea Research Foundation, KOSEF, and Ministry of Education
BSRI 97-2468.

\end{document}